\documentclass[useAMS,usenatbib,usegraphicx]{mn2e}
\usepackage{deluxetable}
\usepackage{epsf}
\usepackage{setspace}

\newcommand{\lya}{Ly$\alpha$\ }

\newcommand{\vw}{{v_{\rm wind}}}

\newcommand{\gad}{{\sc Gadget-2}}
\newcommand{\ion}[2]{\hbox{#1\,{\sc #2}}}
\newcommand{\nhi}{N_{\rm HI}}

\newcommand\cdunits{{\rm cm}^{-2}}

\title{Galactic outflows and the kinematics of damped Lyman alpha absorbers}

\begin{document}

\author[Hong et al.]{
\parbox[t]{\textwidth}{\vspace{-1cm}
Sungryong Hong$^1$, Neal Katz$^1$, Romeel Dav\'e$^2$,
Mark Fardal$^1$, Du\v{s}an Kere\v{s}$^{3}$, Benjamin D. Oppenheimer$^4$}
\\ \\$^1$Astronomy Department, University of Massachusetts, Amherst, MA 01003
\\$^2$Astronomy Department, University of Arizona, Tucson, MA 85721
\\$^3$ Harvard-Smithsonian Center for Astrophysics, Cambridge, MA 02138, USA
\\$^4$ Leiden Observatory, Leiden University, PO Box 9513, 2300 RA Leiden, the Netherlands
}

\maketitle

 \begin{abstract}
The kinematics of damped Lyman-$\alpha$ absorbers (DLAs) are difficult
to reproduce in hierarchical galaxy formation models, particularly
the preponderance of wide systems.  We investigate DLA kinematics
at $z=3$ using high-resolution cosmological hydrodynamical simulations
that include a heuristic model for galactic outflows.  Without
outflows, our simulations fail to yield enough wide DLAs, as in
previous studies.  With outflows, predicted DLA kinematics are in
much better agreement with observations.  Comparing two outflow
models, we find that a model based on momentum-driven wind scalings
provides the best match to the observed DLA kinematic statistics
of Prochaska \& Wolfe.  In this model, DLAs typically arise a few
kpc away from galaxies that would be identified in emission.  Narrow
DLAs can arise from any halo and galaxy mass, but wide ones only
arise in halos with mass $\ga 10^{11}M_\odot$, from either large
central or small satellite galaxies.  This implies that the success
of this outflow model originates from being most efficient at pushing
gas out from small satellite galaxies living in larger halos.  This
increases the cross-section for large halos relative to smaller
ones, thereby yielding wider kinematics.  Our simulations do not
include radiative transfer effects or detailed metal tracking, and
outflows are modeled heuristically, but they strongly suggest that
galactic outflows are central to understanding DLA kinematics.  An
interesting consequence is that DLA kinematics may place constraints
on the nature and efficiency of gas ejection from high-$z$ galaxies.
\end{abstract}


\section{Introduction}

Damped Lyman alpha systems (DLAs), i.e. \ion{H}{i} absorption line
systems having column densities of $\nhi>2\times
10^{20}\cdunits$~\citep{wol86}, contain the dominant reservoir of
neutral hydrogen in the Universe~\citep[see review by][]{wol05}.
Since neutral hydrogen is closely connected with star
formation~\citep[e.g.][]{ken98}, DLAs represent the neutral gas
repository for fuelling star formation.  Influential work by
\citet{sto00} suggested that the global neutral gas content measured
from DLAs declines with cosmic epoch in concert with the growth in
cosmic stellar mass.  Other studies with different selection methods
suggest that the decline in neutral gas is not as steep~\citep{rao06}.
Nevertheless, one expects some connection between galaxies identified
in stellar emission and those identified in gas absorption such as
DLAs.  Studying this connection has been the focus of a large number
of investigations, enabled by the ability to routinely catalogue
absorbers and galaxies at $z\ga 2$ where optical \lya\ absorption
line studies have accurately characterised the DLA population.

Despite these efforts, a deep understanding of the relationship
between DLAs and emission-selected galaxies remains elusive.  Studies
of neutral hydrogen absorption around Lyman break-selected galaxies
(LBGs) at $z\sim 2-3$ show a wide range of absorption
strengths~\citep{ade05}.  DLAs have gas-phase metallicities that
are lower than that seen in LBGs, and show a much larger
spread~\citep{pro05}.  Measures of the \ion{C}{ii}$^*$ cooling rates
in DLAs indicate that the star formation rates are dramatically
lower than that seen in LBGs~\citep{wol04}.  Low-redshift imaging
of DLAs show a heterogeneous population of generally sub-$L_*$
galaxies, but it is unclear if this is relevant to high-$z$ DLAs.
On the other hand, the clustering of LBGs and DLAs at $z\sim 2-3$
suggests that they occupy similar halos~\citep{bou05,coo06}.  Hence,
even though a wealth of observational data exists for DLAs, it
cannot be definitively said whether DLAs are lower-mass LBG analogs,
whether they are a phase of galaxy evolution that eventually leads
to LBGs, or whether they are an altogether separate galaxy population.

One approach for constructing a unified framework for the nature
of DLAs is to employ numerical simulations that account for the
hierarchical growth of galaxies including gas physical processes
and star formation.  Early investigations using simulations by
\citet[hereafter HSR98]{hae98} suggested that DLAs traced clumps of cold gas that
were in the process of assembling into a larger galaxies.  This
proto-galactic clump (PGC) model favoured the interpretation
of DLAs as a precursor phase to LBGs.  \citet{gar01} used simulations
in cosmological volumes to argue that the bulk of DLAs must arise
near dwarf galaxies, suggesting that they are low-mass LBG analogs.
\citet{mal01} used semi-analytic models to argue that DLAs must
come from gas distributions within dark halos that are extended
relative to that expected from angular momentum support, and hence
another mechanism such as tidal stripping may be important for
obtaining the correct cross-section for DLA absorption.  \citet{nag07}
found that DLA properties are significantly affected by galactic
outflows, and argued that the more successful models showed DLAs
as being lower-mass analogs of LBGs.  Simulations by \citet{raz06}
incorporated radiative transfer to model the neutral gas distribution
more accurately, and found that DLAs can arise from a variety of
environments, from the centres of small galaxies to filamentary
tidal structure in large halos.  \citet{pon08} used hydrodynamic
simulations with star formation and outflows together with radiative
transfer, and were able to match the column densities and (for the
first time) the metallicity range of DLAs.  While these successes
are impressive, one set of observations have consistently confounded
all such models:  The kinematic properties of DLAs.

\citet[][hereafter PW97]{pro97} observed the detailed kinematics
of DLAs via their low-ionization metal lines.  They devised four
kinematic diagnostics to quantify their findings, the key one being
the velocity extent of DLAs.  The observations show a pronounced
tail to large velocity extents.  Such a tail is not seen in
simulations, from the earliest models~\citep{pro01} to the most
sophisticated recent ones~\citep{raz08,pon08}.  In general, to match
the observed kinematics the absorption cross section of low-mass
halos must be reduced significantly~\citep{bar09}, which does not
arise naturally in fully dynamical models.

Fundamentally, the difficulty with all these models is that they
yield few absorbers whose velocity spread significantly exceeds the
characteristic velocity of the halo in which the system resides.
Since early gas-rich galaxies tend to be small, this makes it
difficult to reproduce the observed high-velocity ($\ga 200$~km/s)
tail in DLA system widths.  This implies that the gravitational
dynamics of hierarchical assembly alone cannot explain the distribution
of the observed kinematics.  Hence, any model that does not add a
significant component of non-gravitational velocities to the neutral
gas seems doomed to fail the DLA kinematics test.  Alternatively,
the original model of \citet{wol86}, which proposed that DLAs are
large, puffy rotating disks, provides an excellent agreement with
the observed kinematics, but forming enough such objects by $z\sim
3$ is challenging in currently-favoured cosmologies.  Hence the
oft-debated claim by \citet{pro97,pro01} that cold dark matter (CDM)
models cannot straightforwardly reproduce the kinematic structure
of DLAs remains viable.

While it could be that the currently-favoured hierarchical galaxy
formation scenarios are incorrect, their wide-ranging successes in
other areas suggest that the explanation probably lies somewhere
within the poorly-understood baryonic physics associated with galaxy
formation.  For DLAs, it has been suggested~\citep[e.g.][]{pon08}
that galactic outflows may be the missing ingredient needed to
increase the kinematic widths and to lower the absorption cross-sections
of smaller halos.  Recent work indicates that essentially all
star-forming galaxies at $z\ga 1$ are generating outflows of hundreds
of km/s~\citep[e.g.][]{pet01,ste04,sha05,wei09}, with mass outflow
rates comparable to or more than their star formation
rates~\citep{erb06,ste10}.  Theoretically, outflows are believed
to be responsible for suppressing early star formation~\citep{spr03b},
without which the cosmic stellar mass density would far exceed
observations~\citep{dav01,bal01}.  But it is unclear how such
outflows affect DLAs.  Their star formation surface densities are
quite low~\citep{wol04}, so they may not generate outflows at all.
Even if they do, it is unclear how much mass is being carried in
outflows, and whether it would result in enhanced DLA kinematic
widths in accord with observations.  Simulating the impact of
outflows on DLA kinematics is a challenging problem, because it
requires both redistributing neutral gas accurately as well as
self-consistently modeling the energy deposition by outflows into
the surrounding gas.

Recently, several groups have considered the impact of outflows on DLA
properties, specifically kinematics.  \citet{bar09} used a semi-analytic
model to interpret observations of DLAs and faint extended \lya\
emitters, and showed that the incidence rate and kinematics of DLAs
are best reproduced if neutral gas is removed from halos with circular
velocities $\la 50-70$~km/s, suggesting an important role for outflows.
\citet{tes09} followed up the work of \citet{nag07} to assess, among other
things, the impact of various outflow models on DLA kinematics, and found
that none of their outflow models were able to reproduce the kinematics
completely, although a model based on momentum-driven wind scalings came
closest.  \citet{zwa08} argued from an observational perspective that DLA
kinematics must be influenced by outflows, by comparing the \ion{H}{i}
kinematic widths of local \ion{H}{i} galaxies and those of high-$z$
DLAs, and showing that (assuming that locally-identified DLA galaxies are
representative of those at high-$z$) some other process such as outflows
must be enhancing the kinematics of high-$z$ DLAs.  Hence there is growing
evidence that DLA kinematics are strongly impacted by galactic outflows.

In this paper, we investigate whether galactic outflows can both
lower the absorption cross-section of small halos and broaden the
velocity widths enough to bring the theoretical DLA kinematics into
accord with the observations.  To do so, we employ cosmological
hydrodynamic simulations of galaxy formation, with the key addition
being several heuristic models of galactic feedback.  We consider
three variants: No outflows; a model with constant outflow speed
and mass loading factor \citep[i.e.  the mass outflow rate relative
to the star formation rate;][hereafter SH03]{spr03b}; and a model
where the outflow speed and mass loading factor scale according to
expectations from momentum-driven winds~\citep{mur05}.  Such
momentum-driven wind scalings implemented within our models have
proved remarkably successful at reproducing a wide range of
observations, including: the observed IGM metal content from $z\sim
0-6$~\citep{opp06,opp08,opp09}; observations of early galaxy
luminosity functions and evolution~\citep{dav06,bou07}; the galaxy
mass-metallicity relationship~\citep{fin08}; the observed enrichment
in various baryonic phases~\citep{dav07}; the metal and entropy
content of low-redshift galaxy groups~\citep{dav08}; the \ion{H}{i}
distribution in the local universe~\citep{pop09}; and the faint-end
slope of the present day stellar mass function~\citep{opp10}.
Momentum-driven wind scalings of outflow velocity are also observed
in nearby starburst outflows~\citep{mar05,rup05}, providing an
intriguing connection between outflows today and in the past.  Hence
our momentum-driven wind model provides a concrete (albeit
phenomenological) model for how outflows are related to the properties
of star-forming galaxies across cosmic time.

Our primary result in this paper is that our momentum-driven wind
simulation is able to reproduce the observed kinematics of DLAs.
Without winds, our simulated DLA kinematics do not show nearly
enough wide-separation systems and have an average velocity width
that is too small, in agreement with previous studies.  A constant
wind model, as implemented by \citet{spr03b} and used by \citet{nag07}
to study DLAs, improves over the no-wind case but is still unable
to statistically match DLA kinematics.  The primary reason for the
success of momentum-driven wind scalings is that it ejects large
amounts of mass from small galaxies, at relatively low velocities
such that the gas is not overly heated.  Although our simulations
are somewhat simplistic, using a heuristic wind prescription, no
radiative transfer, and relatively low spatial resolution ($\sim200$~pc
physical at $z=3$), they are the first to statistically match the observed 
DLA kinematics within a cosmological context.  The implication is
that DLAs represent the extended neutral gas envelopes of early
star-forming galaxies that are driving outflows, and furthermore
that such outflows are effective at moving large amounts of neutral
gas into the regions surrounding early galaxies.

Our paper is organised as follows:  In \S\ref{sec:sims} we describe
our simulations and methodology for computing DLA absorption.
In \S\ref{sec:kinematics}, we present our results for DLA abundances
and present statistical tests on the kinematics of each feedback model
compared to the observations. We also present many physical properties
that relate DLAs to their host galaxies and dark matter halos.  
We summarise our results in \S 4.

\section{Simulations}\label{sec:sims}

\subsection{Input physics}

We simulate $8~h^{-1}$Mpc periodic cubic comoving volumes using the
cosmological tree-particle mesh-smoothed particle hydrodynamics
(SPH) code \gad, with modifications including radiative cooling and
star formation~\citep{spr03a}.  Our version used in this
work~\citep[described in][]{opp06} also includes metal-line
cooling~\citep{sut93}, assumes a (spatially-uniform) cosmic
photoionising background taken from~\citet{haa01}, and includes a
heuristic prescription for galactic outflows driven by star formation
that we describe in the next section.  We focus on $z=3$ outputs
as that is the typical redshift for the DLAs with well-measured
kinematics.

We choose cosmological parameters of $\Omega_m=0.3, ~\Omega_{\Lambda}=0.7,
~h=0.7,~\sigma_8=0.9,$ and $\Omega_b=0.04$ and generate the initial
conditions at $z=199$ using the \citet{eis99} transfer function.
These parameters are consistent with the WMAP-1 results~\citep{spe03}
but are somewhat different than the latest WMAP-7 results~\citep{kom08}.
However, these differences are not expected to affect our general
conclusions, as the uncertainties in modeling the baryonic physics
probably dominate over these differences.  We use $256^3$ particles
each for the gas and the dark matter, yielding particle masses of
$4.84\times10^{5} M_{\odot}$ for the gas and $3.15\times10^{6}
M_{\odot}$ for the dark matter.  The equivalent Plummer gravitational
softening length is $0.625~h^{-1}$ comoving kpc, or 223~proper pc
at $z=3$.  These simulations resolve star formation in halos down
to virial temperature of $\sim 2\times 10^4$K, below which ambient
photoionisation is expected to provide significant suppression of
gas accretion.  We note that the numerical resolution in our
cosmological volume is better than that in the individual halo
simulations of HSR98.

\subsection{Outflow models}\label{sec:winds}

Galactic outflows appear to be related to star formation, but the
exact relationship is unknown.  While many recent studies have
focused on explicitly driving winds by over-pressurising the ISM
via supernova heat input~\citep[e.g.][]{sti06,cev08}, it is clear
that such processes are sensitive to scales below any realistically
achievable resolution within a cosmological volume.  Hence SH03
took the approach of explicitly incorporating outflows with tunable
parameters, to avoid a strong dependence on physics below the
resolution scale.

In the SH03 implementation, an ``outflow model" is described by
choosing two parameters: the mass loading factor $\eta$, and outflow
velocity $\vw$.  The mass loading factor is defined as $\eta\equiv
\dot{M}_w/\dot{M}_\star$, where $\dot{M}_{w}$ is the mass loss rate
by winds and $\dot{M}_{\star}$ is the star formation rate.

Each gas particle that is sufficiently dense to allow star formation
has some probability to either spawn a star particle, or to be kicked
into an outflow.  The ratio of those probabilities is given by the mass
loading factor $\eta$.  If a particle is selected to be in an outflow,
its velocity is augmented by $\vw$, in a direction given by $\pm${\bf
v}$\times${\bf a} (resulting in a quasi-bipolar outflow).  

When a gas particle becomes a wind particle, its hydrodynamic forces
are turned off until the particle escapes the star-forming region and
reaches a density of one-tenth the critical density for star formation.
The maximum time for hydrodynamic decoupling is 20~kpc divided by the
wind speed, ensuring that the hydrodynamic forces will almost always
be turned off for less than 20~kpc.  At the simulation resolution
we employ, our results are not very sensitive to whether or not we
decouple~\citep[but see][who show that it makes a large difference at
very high resolution]{dal08}, but it does allow for better resolution
convergence.  However, this decoupling does imply that the detailed
distribution of metals around galaxies may be inexactly modeled owing to
the lack of accounting for hydrodynamical effects.  For this reason, we
will prefer to use the \ion{H}{i} distribution directly when computing
DLA kinematics, and simply tie the metals directly to the \ion{H}{i}.
We note that \citet{tes09}, using simulations similar in physics and
resolution to ours, explored the difference in kinematics between using
the metals directly and using the \ion{H}{i} distribution with a fixed
assumed metallicity, and found that observed kinematics were better
reproduced in the latter case, suggesting that some metal diffusion or
smoothing may be required.  We follow their more optimistic case here,
with the hope that future simulations will allow us to more robustly
and self-consistently model the metal distribution near galaxies.

SH03 chose $\vw$ and $\eta$ to be constants, based on observations
available at the time~\citep{mar99,hec00}.  The wind velocity $v_w$
was derived from the supernova feedback energy,
\begin{equation}
\frac{1}{2} \dot{M}_{w} v_w^2 =\chi \epsilon_{SN} \dot{M}_{\star},
\end{equation}
where $\epsilon_{SN}$ is the energy deposition by Type II supernovae
per unit of star forming mass, and $\chi$ is the fraction of
supernovae energy that drives the wind.  SH03 set $\chi=1$, and
used $\epsilon_{SN}\sim 4\times 10^{48}$~erg$/M_\odot$ for a Salpeter
initial mass function, yielding a wind velocity of $\vw=484$~km/s.
They set $\eta=2$ to roughly reproduce the observed
present-day mass density in stars.  SH03 showed that this wind model
produces a cosmic star formation history that is in broadly agreement with
observations, and is resolution converged.  We will refer to this
as the constant wind (``cw") model, since both $\eta$ and $\vw$ are
independent of galaxy size.

Improved recent observations of local starburst outflows suggest
that the wind speed is not constant, but is proportional to a
galaxy's circular velocity~\citep{mar05,rup05,wei09}.  This is
suggestive of momentum-driven winds, as worked out analytically by
\citet{mur05}.  OD06 implement a momentum-driven wind model in a
cosmological simulation.  In this scenario, the wind speed scales
as the galaxy's velocity dispersion $\sigma$, and since the momentum
deposition per unit star forming mass is assumed to be constant,
the mass loading factor is inversely proportional to $\sigma$.

For our momentum-driven wind model, we employ
\begin{eqnarray} 
v_w &=& 3\sigma\sqrt{f_L -1}\\ 
\eta &=& \frac{\sigma_0}{\sigma} ,
\end{eqnarray} 
where $f_L$ is the galaxy
luminosity divided by its Eddington luminosity and $\sigma_0$ is a
normalisation factor.  Because our resolution is still too low to compute
the galaxy's velocity dispersion, we take $\sigma = \sqrt{-\frac{1}{2}
\Phi}$ from the virial theorem, where $\Phi$ is the local potential depth
where a wind particle is created.  The normalisation factor $\sigma_0$
is set to 300~km/s as suggested by \citet{mur05}.  Values for $f_L$ are
taken from observations: \citet{mar05} found $f_L\sim 2$, and 
\citet{rup05} measured $f_{L} \approx 1.05 - 2$, for local
starbursts.  In the momentum-driven wind scenario, lower metallicity
star formation should produce stronger outflows owing to greater
ultraviolet (UV) flux output; we account for this using the stellar
models of \citet{sch03}.  Specifically, we set
\begin{equation} 
f_L = f_{L\odot} \times 10^{-0.0029\times(\log Z + 9)^{2.5}+0.417694} 
\end{equation} 
where $f_{L,\odot}$ is randomly chosen between $1.05-2.0$, and $Z$ is
the gas particle's metallicity.  We term this momentum-driven wind model
(``vzw").  In contrast with the cw model, $\vw$ and $\eta$ depend on
galaxy size.

We note that this momentum-driven wind implementation differs from
our more recent works~\citep[e.g.][]{opp09} that use the galaxy mass
to calculate $\sigma$.  However, as we showed in~\citet{opp08}, this
choice makes little difference at $z=3$, and only becomes critical at
$z\la 2$ when large potential wells with hot gas develop.  Furthermore,
our implementation also differs from that in \citet{tes09}, who use a
calibrated relationship to the
halo mass to determine $\sigma$.

Finally, for comparison we also evolve a model with no galactic
outflows (no winds, ``nw").  Note that this model still includes
thermal supernova feedback within the context of the subgrid
multiphase ISM model of \citet{spr03a}.  It is a poor match to many
observations, but provides a baseline to assess the impact of
outflows.  The suite of three models employed here are a subset of
those in OD06.

\subsection{Distribution of neutral gas}

Here we describe how we calculate the distribution of neutral gas
in each simulation.  There are two significant complications for
this: first, in the multi-phase ISM model of \citet{spr03a} that
we employ, only some fraction of each particle is actually neutral.
Second, since we do not perform detailed radiative transfer, we
must account for the effects of self-shielding that can mitigate
the strength of the photoionising flux.

The first issue is straightforward to handle, at least to a sufficient
approximation.  In our multi-phase model, each particle above a
critical density for fragmentation is assumed to consist of a cold
phase at 1000~K plus a hot phase at $10^6$~K.  The cold phase
dominates the mass fraction, ranging between $84-100$\% (for
primordial gas).  Since we are only interested in the cold phase,
we assume all particles above the fragmentation density to have
90\% of their mass in the cold phase at 1000~K, and ignore the hot
phase.  Note that with metal cooling included, the fragmentation
scale can depend on particle metallicity~(OD06) and the cold mass
fraction can range slightly lower, but to the level of approximation
required here this is not important.  In detail, the cold phase is
probably confined to even smaller sub-particle scale clumps, but
as long as the overall cross-section of these clumps is close to
unity on galactic scales \citep[as is found observationally when
examining the \ion{H}{i} content of galaxy disks, e.g.][]{zwa05},
then the assumption of smoothly-distribution gas within the particle
seems reasonably valid.

Accounting for self-shielding requires more significant approximations.
The observed HI column density distribution shows an excess of
absorbing systems with $N(HI) \geq 2\times 10^{20}$ when the
distribution is extrapolated from the low column density region
populated by the Lyman $\alpha$ forest~\citep{lan91,sto00}.  This
excess has been explained as a self-shielding effect typically
causing a sharp transition layer from mostly ionised regions to
mostly neutral regions~\citep{mur90,pet92,cor01}.  Hence self-shielding
is a critical aspect for setting the ionization level of gas in
DLAs.

Owing to our lack of a full radiative line transfer treatment of
\lya\ photons, we resort to a simpler estimate for self-shielding.
HSR98 use a pure density criterion which assumed that all the gas
with a number density above 0.01~cm$^{2}$ is fully self-shielded,
while the rest is subject to the full metagalactic UV flux.  In
reality, such a threshold depends on the size of the collapsed
object and the local strength of UV background, and hence should
be different at each local position.

Our approach is to have two criteria to identify a gas particle as
``self-shielded": a maximum temperature and a minimum density. We
choose a single value for each criterion within each wind model,
and use the observed abundance of DLAs to constrain this value.  We
will demonstrate that the DLA abundances are not sensitive the
choice of temperature threshold, and that the DLA kinematics obtained
using different (reasonable) thresholds do not differ significantly.
For densities, we will employ comoving densities of
$\rho_\theta=(40,80)\rho_{\rm crit}$, where $\rho_{\rm crit}$ is
the critical density at $z=0$, as two thresholds that span a
reasonable range.  These correspond to $(1000,2000)\bar\rho_b$, or
a baryon density of approximately $(0.014,0.028)~$cm$^{-3}$ (physical).

In Table \ref{table:names} we summarise our models.  The names
signify the wind model and $\rho_\theta$.  Also shown are the DLA
number densities per unit redshift, and the number of randomly
chosen lines of sight (LOS) in our DLA sample.  These LOS will be
used to investigate the kinematics of each model.  Models in boldface,
cw40 and vzw80, are the ones that broadly match the observed number
density of DLAs, as we will discuss further in \S\ref{sec:dndz},
while also having an observationally-consistent cosmic star formation
rate at $z\sim 3$.  When we need to compare these wind models with
the no-wind case, we choose nw80 as a representative, although this
model produces too many DLAs and strongly overpredicts the cosmic
star formation rate.  Note that in order to bring the no-wind model
into agreement with the number density of DLAs, we need a very high
choice of $\rho_\theta$ that is likely to be unphysical (in the
sense that this density well exceeds the threshold density for our
multi-phase ISM model).  This further highlights the impact that
outflows have in regulating the amount of neutral gas in DLAs and
the cosmos.

\begin{deluxetable}{lccccc}
\footnotesize
\tablecaption{}
\tablewidth{0pt}
\tablehead{
\colhead{Name} &
\colhead{$\rho_\theta$\tablenotemark{a}} &
\colhead{$dN/dz$\tablenotemark{b}} &
\colhead{$N_{\rm LOS}$\tablenotemark{c}} &
\colhead{$N_{\rm 20 km/s}$\tablenotemark{d}} &
\colhead{$N_{\rm 30 km/s}$\tablenotemark{e}}
}
\startdata
nw40  & 40 & 0.534 & 1995 & 883 & 404\\
nw80  & 80 & 0.443 & 1995 & 716 & 310\\
{\bf cw40}  & 40 & 0.244 & 1995 & 1199 & 672\\
vzw40  & 40 & 0.379 & 1992 & 1306 & 973\\
{\bf vzw80}  & 80 & 0.265 & 1988 & 1453 & 1175\\
\enddata
\tablenotetext{a}{Self-shielding comoving density in units of the $z=0$ critical density.}
\tablenotetext{b}{Number of DLAs per unit redshift.}
\tablenotetext{c}{Number of lines of sight.}
\tablenotetext{d}{\# of DLAs with $\Delta v>20$~km/s and satisfying eq.~\ref{eqn:Ipk}.}
\tablenotetext{e}{\# of DLAs with $\Delta v>30$~km/s and satisfying eq.~\ref{eqn:Ipk}.}
\label{table:names}
\end{deluxetable}

\subsection{DLA absorption profiles}\label{sec:profiles}

We now identify DLAs and calculate the absorption profiles for
low-ionisation species.  We first perform self-shielding corrections
to the neutral gas content of individual gas particles as described
in the previous section.  To identify DLAs, we make an 8000$\times$8000
\ion{H}{i} map projected through the entire simulation volume.  We
randomly choose 2000 pixels with $\nhi>10^{20.3}\cdunits$, and
compute \ion{H}{i} \lya\ spectra along lines of sight (LOS).  We
then identify the individual absorbing region along the LOS that
is a DLA.  In less than 1\% of the cases, the total $\nhi$ of any
single absorber along the LOS is below the DLA threshold, even
though the total column exceeds it; we discard those LOS.  The
actual numbers of identified DLAs for each model are listed in
Table~1.

To obtain DLA kinematics, we attempt to mimic the observational
procedure of identifying low-ionisation, relatively unsaturated
metal lines that can trace the gas motions.  Although we track metal
enrichment in our simulations, we choose not to attempt to make
spectra from the metals directly.  This is because (1) we don't
accurately track the ionisation field, and (2) as explained above,
the tracking of metals in outflows near galaxies is hampered by
limited numerical resolution and explicit hydrodynamic decoupling.
Instead, we tie the metal abundance to the \ion{H}{i} abundance
with values representative of fully neutral gas.

We choose \ion{Si}{ii} (1808\AA) as our representative low-ionisation
metal tracer, which was used in HSR98 and is generally a good tracer
for DLA kinematics.  We assume a uniform silicon abundance of
[Si/H]$=-1$, and take the oscillator strength from \citet{tri96}.
Since the ionisation potentials of \ion{Si}{i} and \ion{Si}{ii} are
8.15 eV and 16.3 eV, respectively, which brackets \ion{H}{i}, we
assume that SiII is dominant and exists only in self-shielded regions
and in cold ISM gas.  This gives an abundance of $n_{\rm SiII}/n_{\rm
HI}=3.24\times 10^{-6}$.  We multiply the \ion{H}{i} optical depths
along each spectrum by this value, along with the oscillator strength
difference, to obtain the \ion{Si}{ii} optical depths, thereby
generating spectra from which we can examine DLA kinematics.  Our
results are not sensitive to the specific abundance and ionization
choice, because we are primarily interested in the redshift-space
extent of metal systems rather than the strengths of the systems
themselves.

\begin{figure}
\vskip -0.15in
\setlength{\epsfxsize}{0.5\textwidth}
\centerline{\epsfbox{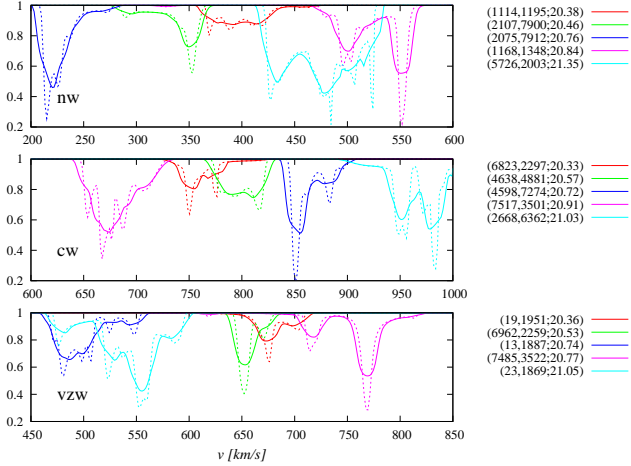}}
\caption{Example low-ionisation metal line (i.e. \ion{Si}{ii}) 
spectra drawn from our simulations.  The metal optical depths are
tied to \ion{H}{i} as described in the text.  Each colour shows a 
different LOS; solid and dotted lines show smoothed (19~km/s tophat) and 
unsmoothed versions, respectively.
}\label{fig:profiles}
\end{figure}

Example randomly-chosen spectra are shown in Figure~\ref{fig:profiles}.
Each panel contains 5 LOS; the locations (in our $8000\times 8000$ grid)
and \ion{H}{i} column densities are indicated along the right side.
Such spectra are analyzed as described in \S\ref{sec:stats} in order to
obtain kinematic statistics for each DLA.

\subsection{Identification of galaxies and dark matter halos}

Besides DLA kinematics, we are also interested in studying the
relationship between DLAs and galaxies in our simulations.  Hence,
we must identify galaxies in our simulations and we do so using
SKID\footnote{Spline Kernel Interpolative DENMAX;
http://www-hpcc.astro.washington.edu/ tools/skid.html}~\citep{kat96a},
which identifies density watersheds using spline kernel interpolation,
and then finds surrounding groups of gas and star particles that
are gravitationally self-bound within these density watersheds.  We
apply SKID to all the star particles and gas particles that satisfy
$\rho_{\rm gas}/\bar{\rho}_{\rm gas} > 10^3$ and $T < 30,000$K, and
all particles (regardless of temperature) above the multiphase
threshold of $\approx 1000\bar{\rho}_{gas}$.

\begin{figure}
\vskip 0.1in
\setlength{\epsfxsize}{0.5\textwidth}
\centerline{\epsfbox{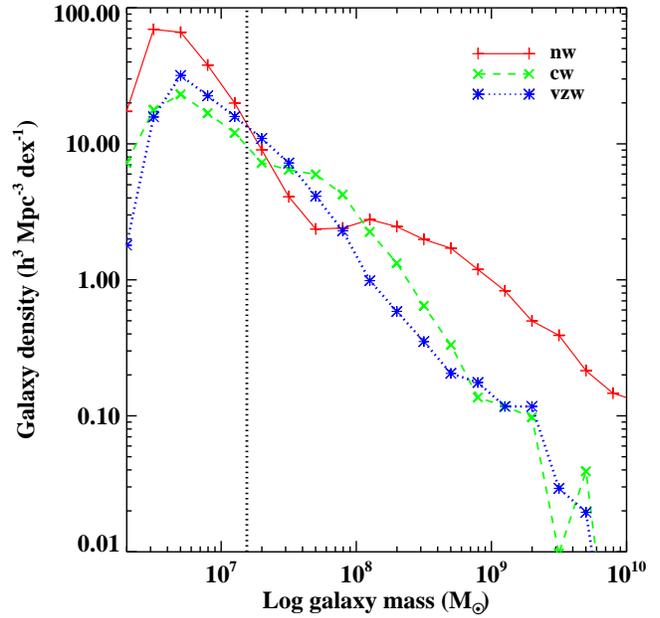}}
\vskip -0.15in
\caption{The mass function of gravitationally bound groups found by SKID 
for the three models: no winds (nw, red solid), constant winds (cw, green
dashed) and momentum-driven winds (vzw, blue dotted).
We resolve galaxies down to a baryonic mass of $1.55\times10^{7} M_{\odot}$,
indicated by the vertical dotted line.
}\label{fig:galmass}
\end{figure}

Figure~\ref{fig:galmass} shows the galaxy baryonic mass function
for our three simulations We take our galaxy mass resolution limit
to be $32 m_{\rm SPH} = 1.55\times 10^{7} M_{\odot}$~\citep{fin06},
shown as the vertical line.  This limit appears to be conservative,
as the mass function doesn't turn over severely until much lower
masses, but detailed resolution testing shows that the individual
galaxy star formation histories are poorly converged below this
mass owing to stochastic effects and the fact that the densities
cannot be adequately resolved.  The inflection point in the no-wind
case we believe corresponds to the filtering mass~\citep[e.g.][]{gne00},
which is the mass around which galaxies are significantly suppressed
by metagalactic photoionization.  Galaxies below this mass formed
prior to turning on our spatially-uniform ionizing background
instantaneously at $z=9$~\citep{haa01}.  A more realistic radiative
transfer simulation would not yield such instantaneous reionization,
and would likely not produce such a feature.  The wind models have
much lower stellar masses for a given halo mass, and hence the
effects of filtering are relegated to much smaller masses, at or
near our resolution limit.  While these effects on the mass function
are interesting, they are not immediately relevant to DLAs, so we
do not discuss them further.

To identify dark matter halos we use a friends-of-friends(FOF)
algorithm~\citep{dav85}, which finds groups of dark matter particles
by linking neighbouring particles within a given linking length.
We choose the linking length to be the interparticle distance at
one-third of the virial overdensity $\rho_{vir}/\bar{\rho}$, which
is the local density at the virial radius for an isothermal sphere,
and we take $\rho_{vir}$ from \citet{kit96}.  To define the virial
mass, virial radius, and the circular velocity of the dark matter
halo we refine the group using a spherical overdensity (SO) criterion.
In SO, the halo centre is set to be the location of the most bound
FOF particle. Then we expand the radius around this centre until
the mean overdensity inside the radius equals $\rho_{\rm vir}/\bar{\rho}$.
We define this radius to be $R_{\rm vir}$, the mass within
$R_{\rm vir}$ to be $M_{\rm vir}$, and the circular
velocity $v_c = \sqrt{GM_{\rm vir}/R_{\rm vir}}$.

In the end, we obtain a sample of about 2000 DLAs and $\sim 3000$
galaxies within each of our simulations.  We will now analyze these
systems to understand the impact of outflows on DLA kinematics and
the relationship between DLAs and galaxies.
  
\section{Physical properties of DLAs}\label{sec:kinematics}

\subsection{DLA number densities}\label{sec:dndz}

\begin{figure}
\vskip -0.15in
\setlength{\epsfxsize}{0.5\textwidth}
\centerline{\epsfbox{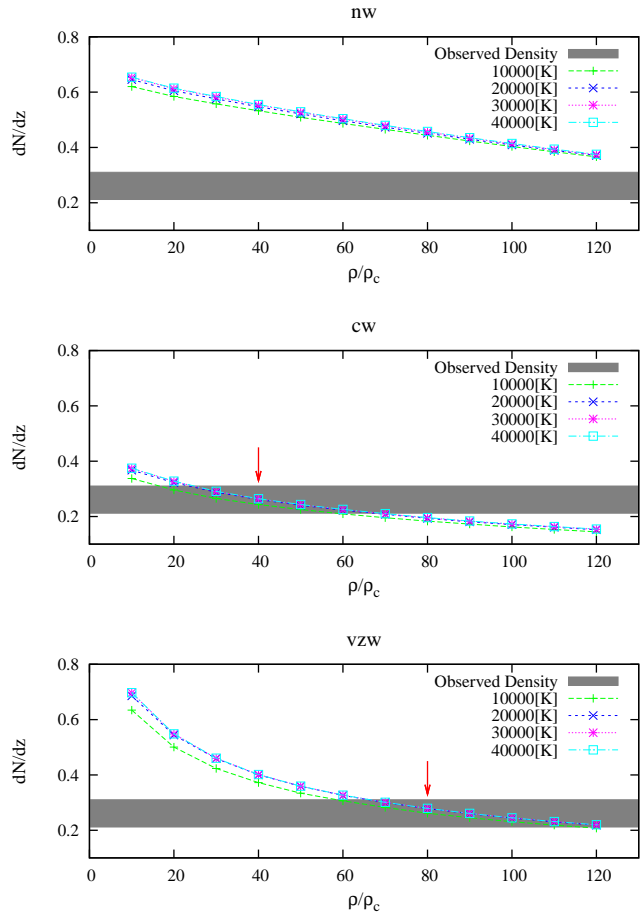}}
\caption{DLA abundances for each model assuming various
self-shielding criteria. The temperature dependence is weak, while the
density dependence is strong. The red arrows indicate the value
we choose for $\rho_\theta$. Note that 
the values in this figure are slightly 
larger than the values in Table \ref{table:names} because 
we used 2000x2000 maps for this figure and 8000x8000 maps for our 
main results as presented in Table \ref{table:names}.
}\label{fig:abundance}
\end{figure}

The observed redshift-space abundance of DLAs, $dN/dz = 0.26\pm
0.05$~\citep[$z=3$;][]{sto00,pro05}, provides a key constraint for DLA
models.  In our case, we use it to constrain our self-shielding criteria,
$\rho_{\theta}$ and $T_{\theta}$.  Figure \ref{fig:abundance} shows our
derived abundance for each wind model assuming different $\rho_{\theta}$
and $T_{\theta}$.  Remember that we assume that all gas with a density
greater than $\rho_{\theta}$ and a temperature less than $T_{\theta}$
is fully self-shielded. Overall, changing $T_{\theta}$ has little effect
on the abundance, while $\rho_{\theta}$ has a large effect.  

The no-wind (nw)  model has too much neutral gas to match the observations
within a large range of choices for $\rho_{\theta}$.  For the two wind
models, we choose two self-shielding criteria, $\rho_{\theta} = 40$ and
$\rho_{\theta} = 80$ for the cw model and the vzw model, respectively, to
approximately match the observed $dN/dz$.  A density of $\rho_{\theta}=40$
corresponds to $1000\bar{\rho}_b$ and $n_H = 0.014\;{\rm cm}^{-3}$ in
primordial composition gas, which was adopted in HSR98 as their shielding
criteria.  So the cw model has the same density threshold value as HSR98
while the vzw model has one twice as large. Our resolution and inclusion
of metal line cooling likely affects the choice of threshold values.

\begin{figure}
\vskip -0.1in
\setlength{\epsfxsize}{0.4\textwidth}
\leftline{\epsfbox{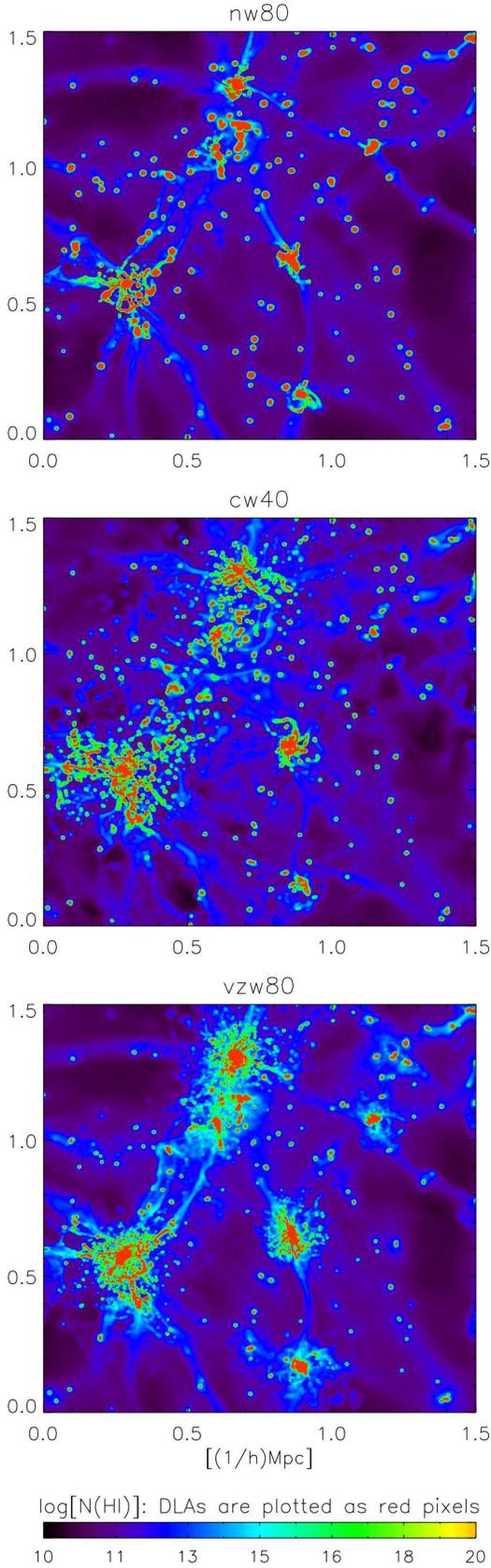}}
\vskip -0.1in
\caption{The column density of neutral hydrogen (HI maps) for the no wind model
(nw80, top panel), the constant wind model (cw40, middle panel) and the
momentum driven wind model (nw80, bottom panel). 
Every pixel over $2\times 10^{20} {\rm cm}^{-2}$, i.e. damped absorption,
is plotted in red.  All lengths are comoving.
}\label{fig:himaps}
\end{figure}

Figure \ref{fig:himaps} shows HI column density maps projected along
the x-axis for the nw80, cw40, and vzw80 models. We see substantial
differences in the neutral gas distributions produced by these three
different feedback implementations.  Without winds, the DLA-absorbing
gas is highly concentrated within star-forming regions.  In the
constant wind (cw) case, the absorption is somewhat more extended,
but the high outflow velocities generally cause the outflowing gas
to be heated, thereby lowering their neutral fractions and hence
their densities.  The momentum-driven wind model (vzw) produces the
greatest extent of high-column density gas.  This foreshadows our
result that the vzw model will yield the largest DLA kinematic
widths.

The more extended gas distribution causes the vzw model to be most
sensitive to $\rho_{\theta}$.  This subtle dependence arises because the
abundance is sensitive to cross-section, since an extended distribution
results in more gas lying near the threshold density.  So the curves in
Figure \ref{fig:abundance} partly reflect the different gas topologies
that result from the different outflow models. In contrast, the
sensitivity to temperature threshold is negligible because most of the
gas for any reasonable choice of $\rho_{\theta}$ lies at $\sim 10^4$~K,
below which we truncate radiative cooling in our simulations.

As an aside, we note that the filamentary gas structures that are
responsible for feeding galaxies with fresh fuel via the cold mode
accretion scenario~\citep[e.g.][]{ker05,dek09} are generally not
sufficiently dense to produce DLA absorption, and instead typically
have $\nhi\la 10^{17}\cdunits$.  Only in the vicinity of galaxies
does the gas starts to self-shield, and the DLAs tend to be confined
to even denser gas within and around galaxies.

\subsection{DLA kinematics}\label{sec:stats}

We now examine the primary target of our investigation: The kinematics of
simulated DLAs and how they compare to observed kinematics.  PW97 developed
a suite of statistics to descibe the one-dimensional distribution of
metal absorption lines within a single DLA.  To generate these statistics
from our simulated DLAs, we adopt the following procedure:
\begin{enumerate}
\item We select DLAs and then generate \ion{Si}{ii} absorption lines by
tying \ion{Si}{ii} to \ion{H}{i} as described in \S\ref{sec:profiles}.\\
\item We smooth the absorption lines with a 19 km/s tophat window, which 
is equivalent to the 9-pixel tophat smoothing done in PW97.  \\
\item From this profile, we
calculate four key quantities, as defined in PW97:  The velocities of the
largest ($v_{\rm pk}$) and second largest peak absorption ($v_{\rm 2pk}$), the median 
absorber velocity
($v_{\rm med}$), and the system velocity width ($\Delta v$) defined
by
\begin{equation}
\Delta v \equiv |v_{95} - v_{5}|,
\end{equation}
where $v_{95}$, $v_5$, and $v_{\rm med}$ represent the positions 
at which the optical depths are 95\%, 5\%, and 50\% of the total optical depth.
We note that we compute these quantities directly from the \ion{Si}{ii}
optical depths, while PW97 compute them from apparent optical depths
obtained from the flux; we expect these quantities to be similar,
particularly for the typically unsatured metal lines in DLAs.

\item From these quantities, we calculate the DLA kinematic statistics from PW97: 
\begin{eqnarray}
f_{\rm mm} = \frac{|v_{\rm med} - v_{\rm mean}|}{(\Delta v/2)} \\
f_{\rm edg} = \frac{|v_{\rm pk} - v_{\rm mean}|}{(\Delta v/2)} \\
f_{\rm 2pk} = \pm\frac{|v_{\rm 2pk} - v_{\rm mean}|}{(\Delta v/2)} 
\end{eqnarray}
where the plus sign for the two peak fraction, $f_{\rm 2pk}$, holds
if the second peak is between the mean velocity and the first peak velocity; 
otherwise the minus sign holds.  
If there is no second peak, we take the edge-leading fraction, 
$f_{\rm edg}$, for the second peak fraction $f_{\rm 2pk}$. 
Here, $v_{\rm mean} \equiv \frac{1}{2}(v_{95}+v_{5})$.

To avoid saturation effects that blur the kinematic information, and to
exclude noise contamination in the observations, we again follow PW97 by
only using profiles with peak intensities $I_{\rm pk}$ in the range
\begin{equation}\label{eqn:Ipk}
0.1 \leq \frac{I_{\rm pk}}{I_0} \leq 0.6
\end{equation}
where $I_0$ is the continuum around the absorption line, or
peak optical depths between 0.5 and 2.3.  This removes a significant
portion of the total DLA sample, roughly 60\% for nw, 45\% for cw, and
30\% for vzw.  The variance amongst models in the acceptance fraction 
based on this criterion suggests that this may be another way to 
constrain outflow models.

\end{enumerate}

The most crucial statistic is $\Delta v$, the system velocity
width.  It represents the velocity-space extent of the dense neutral
absorbing gas, and hence encodes information about internal motions
within the ISM as well as any inflow or outflow-induced motions.
The other statistics, $f_{\rm mm}$, $f_{\rm edg}$, and $f_{\rm
2pk}$, turn out to be less discriminatory, but we include them for
completeness.  $f_{\rm mm}$ measures the skewness of the overall
absorption in the DLA; a symmetric distribution of optical depths
would yield $f_{\rm mm}=0$.  The edge-leading test $f_{\rm edg}$
would be 0 if the strongest absorption is at the kinematic center,
but is large if the kinematics are dominated by rotation or infall
where the strongest absorption occurs at large velocities from the
center.  The 2-peak test $f_{\rm 2pk}$ is designed to distinguish
between rotation and infall: in the case of rotation, the second
peak is expected to be on the same side as the first (and hence
yield a positive value), whereas spherical and symmetric accretion
would produce the peaks on opposite sides, yielding a negative
value~(PW97).  While these statistics were devised to distinguish between
simple scenarios, the complex interplay between infall, outflow,
and rotation within a fully hierarchical context precludes such
straightforward interpretations.  Hence we focus on the distributions
of these statistics among DLA samples (both observed and simulated),
and use Kolmogorov-Smirnoff (K-S) tests to characterize their
(dis)agreement.

In Table \ref{table:names} we summarise the number of simulated
absorption lines, $N_{\rm sample}$, taken from each model.
For the observations, we will compare to 46 observed lines of sight from
\citet{pro01}.  Note that their paper only presented the distribution
of $\Delta v$; the other quantities were kindly provided by X. Prochaska
(private communication).

Before we compare the simulations with the observations, we must
consider the issue of velocity resolution effects in the observed
DLA sample.  The 9 pixel smoothing of PW97 effectively provides a
minimum of $\Delta v_{min}\sim20$~km/s, and indeed all of the
observed DLAs have $\Delta v>20$~km/s.  But HSR98 adopted a minimum
threshold of $\Delta v_{min} = 30$~km/s to avoid any possible
incompleteness of the DLA sample in the range of $\Delta v=20-30$~km/s.
Unfortunately, this threshold can bias the statistics, as it turns
out that there are a significant number of smaller halos that host
DLAs having $\Delta v$ within this range.  To test this incompleteness
issue, we present results for both velocity thresholds of 20~km/s
and 30~km/s.  However, we note that the observed sample may be
significantly incomplete for the 20~km/s threshold, and may not be
a fair comparison to the models, hence we prefer to compare to the
30~km/s case.  As it happens, our main conclusions are unaffected
by this choice.

\begin{figure*}
\vskip -0.6in
\setlength{\epsfxsize}{0.85\textwidth}
\centerline{\epsfbox{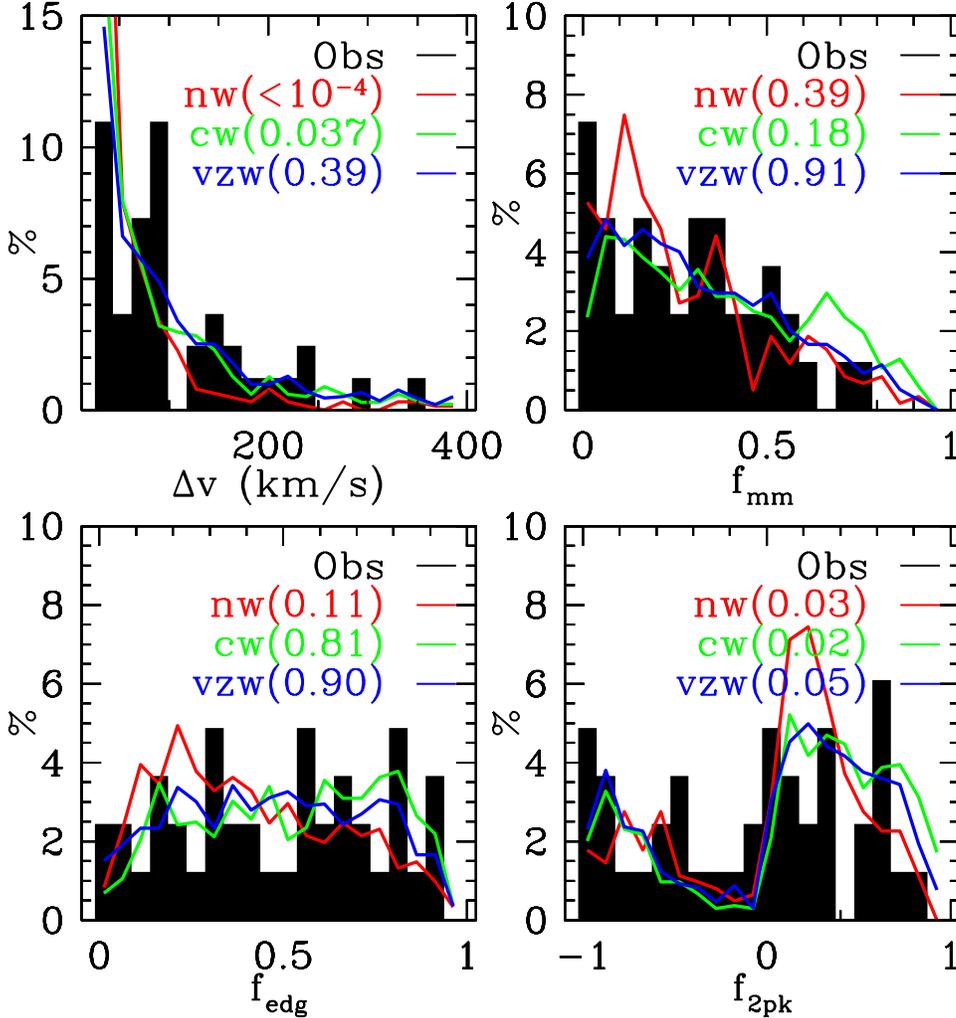}}
\vskip -2.1in
\caption{Distributions of the four statistics from PW97 (shaded histogram)
compared with the nw80 (red), cw40 (green), and vzw80 (blue) models.
A minimum velocity threshold of $30$~km/s is assumed. Numbers in
parenthesis are K-S test values comparing each model with observations,
summarised in Table \ref{table:statistics}.  The no-wind simulations
is strongly disfavored, and statistically the vzw
model provides better agreement than cw.
}\label{fig:showkstest}
\end{figure*}

Figure \ref{fig:showkstest} plots histograms of the four PW97
statistics for our three outflow models, along with their K-S test
values as compared to the observed statistics, with $\Delta v_{\rm
min}=30$ km/s.  The upper left panel demonstrates the central result
of our paper:  Models with outflows provide a much better match to
PW97's velocity width statistic than our no-wind case.  In particular,
the momentum-driven wind simulation provides an excellent match to
the data with a K-S acceptance level of 0.39, although the constant
wind case cannot be definitively ruled out.  The rest of the paper
will mostly be devoted to understanding the physical origin of the
differences among model for the velocity width statistic $\Delta
v$.  We will see that the variations arise because the differences
between wind models are most significant for small halos.

The other PW97 tests generally show acceptable K-S values for all
models when compared to the data, at least for $\Delta v_{\rm
min}=30$~km/s; at $\Delta v_{\rm min}=20$~km/s again the vzw case
is clearly favored as seen in Table~\ref{table:statistics}.  The
median-mean statistic $f_{\rm mm}$ (Figure~\ref{fig:showkstest},
top right) shows that the cw model produces slightly more skewed
DLAs than the vzw and nw cases, resulting in marginally poorer
agreement in the K-S test, but not so significant as to reject the
model.  The edge-leading statistic $f_{\rm edge}$ (bottom left)
tells a similar story, that nw produces the fewest edge-leading
profiles and cw the most, but none of the models can be excluded
by the data.  The two-peak test (bottom right) shows that most
galaxies show rotation signature in their center and interception
with infalling or outflowing gas is rare, but it is more frequent
for the wind models.  This last statistics shows the most disagreement
between the simulations and the data, and none of the models are
obviously favored.  It is unclear why this discrepancy is strongest,
since it is a difficult statistic to interpret.  However, we note
that the thick disk model favoured by PW97 shows a similar
distribution, and hence is expected to have
roughly the same K-S value.

Our no-wind results are significantly different than HSR98, who found
broad agreement with their simulations (without strong outflows) and the
data.  The discrepancy arises because they extrapolate the relationship
between DLA cross-section and halo mass to low halo masses in order to
obtain cosmologically-based statistics for their DLA velocity widths.
In \S\ref{sec:crosssection} we calculate this relationship directly
from our simulations, and show that their extrapolation is not valid.
Specifically, our no-wind simulation shows considerably increased
cross-section over such an extrapolation.  This is the main effect that
causes our no-wind case to be in poor agreement with data, as compared
to HSR98.  We note that our numerical resolution is similar to, in fact
slightly better than, HSR98.  Since HSR98 simulated individual galaxies
while we employ a full cosmological simulation, we are able select
DLAs based upon their absorption cross-sections to match the observed
abundance of DLAs (which HSR98 could not do).  This enables us to more
accurately characterize the simulated DLA population and more robustly
compare it to observations.

HSR98 also noted that their results depended on the assumed
self-shielding criterion.  In Table~\ref{table:statistics} we present
the results for different self-shielding criteria as well as for
different minimum velocity thresholds.  The K-S tests indicate that
the vzw model is consistent with the observations for either choice
of the threshold density, while the no-wind case is strongly
discrepant in $\Delta v$ regardless of threshold choice.  Hence
reasonable variations in the self-shielding criteria do not
significantly impact our overall conclusions.  Hence while it
certainly possible to perform more sophisticated self-shielding
models even without radiative transfer~\citep[e.g.][]{pop09}, or
even do the full post-processing radiative transfer~\citep{pon08},
we do not expect that this will strongly impact our results.

\begin{deluxetable}{lccccc}
\footnotesize
\tablecaption{{K-S results}}
\tablewidth{0pt}
\tablehead{
\colhead{Model} &
\colhead{$v_{\rm min}$} &
\colhead{$P_{\Delta v}$} &
\colhead{$P_{\rm mm}$} &
\colhead{$P_{\rm edge}$} &
\colhead{$P_{\rm tpk}$}
}
\startdata
nw40  & 20 & $<10^{-9}$ & $<10^{-5}$ & $<10^{-6}$ & $<10^{-3}$  \\
nw80  & 20 & $<10^{-11}$ & $<10^{-7}$ & $<10^{-6}$ & $<10^{-3}$  \\
{\bf cw40}  & 20 & $<10^{-6}$ & 0.0036& $<10^{-3}$ & 0.007  \\
vzw40  & 20 & 0.18 & 0.70& 0.58 & 0.14\\
{\bf vzw80}  & 20 & 0.028 & 0.14& 0.18 & 0.08  \\
\hline
nw40  & 30 & $<10^{-4}$ & 0.69 & 0.19 & 0.005  \\
nw80  & 30 & $<10^{-5}$ & 0.39 & 0.11  & 0.03  \\
{\bf cw40}  & 30 & 0.037 & 0.18 & 0.81 & 0.02  \\
vzw40  & 30 & 0.51 & 0.80& 0.92 & 0.04\\
{\bf vzw80}  & 30 & 0.39 & 0.91 & 0.90 & 0.05  \\
\enddata
\label{table:statistics}
\end{deluxetable}

In summary, galactic outflows appear to be capable of reconciling
CDM-based models of galaxy formation with observed DLA kinematics.
This is the most important conclusion from our work.  Such statistics also
provide constraints on wind models, and in the limited tests conducted
here we favour momentum-driven wind scalings over constant wind scalings.
We now investigate these trends more deeply by studying the properties
of galaxies and halos that give rise to DLAs, in order to understand
the physical reasons behind the success of outflows in general, and the
momentum-driven wind scalings in particular.

\subsection{Distance to DLA host galaxies}

In our simulations, DLAs are generally associated with galaxies.  This is
believed to be true in the real Universe, although it is usually only
possible to image the host galaxies at low redshifts, where galaxies can
be separated from the bright background quasar.  On the other hand,
DLAs seem to deviate from well-established properties of galaxies such as
the Kennicutt-Schmidt relation~\citep{wol06}.  Furthermore, \citet{wol08}
identified a bimodality in DLAs, in which they divide into systems with
high and low [CII]~158~$\mu$m cooling rates, with the former having
properties similar to Lyman Break Galaxies and the latter perhaps arising
in a different sort of population.  Hence the relation between DLAs
and galaxies remains uncertain.

Here we examine some basic properties of the relationship between DLAs
and galaxies in our simulations.  To connect DLAs to their host galaxies
and their host dark matter halos, we need to define the line-of-sight
positions of DLAs.  Since it is somewhat arbitrary to define the
position of diffuse gas, we consider two definitions:  $x_{50}\equiv$
the line-of-sight position where 50\% of the optical depth is reached;
and $x_{\rm ave}\equiv 0.5(x_{30}+x_{70})$, where $x_{30}$ and $x_{70}$
are defined analogously to $x_{50}$.  We also define an uncertainty on
the line-of-sight position as $\Delta x\equiv |x_{50} - x_{\rm ave}|$.

In the vast majority of cases, $\Delta x$ is quite small, so the
distance is fairly unambiguous.  The few percent of cases where it
is not are typically caused by two or more separate gas clumps along
the line of sight contributing to a single system.  For the remainder
of this paper, we will exclude cases where $\Delta x>0.01$ in units
of the simulation box length (i.e. 28.7 proper kpc), since these
systems are difficult to relate unambiguously to galaxies.
Specifically, this excludes 1.5\%, 6.9\%, and 3.3\% of DLAs in the
nw, cw, and vzw cases, respectively.  We also examined a tolerance
criterion of $\Delta x=0.001$, with negligible difference in the
final results.

\begin{figure}
\vskip -0.15in
\setlength{\epsfxsize}{0.55\textwidth}
\centerline{\epsfbox{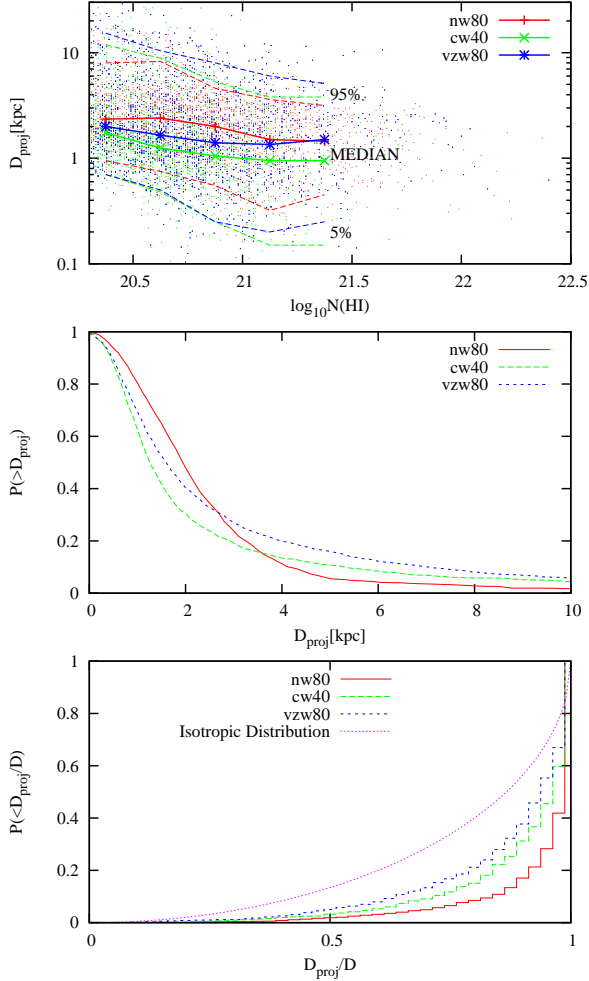}}
\caption{The projected distance, $D_{\rm proj}$, from the host galaxy (top panel),
its cumulative distribution (middle panel), and the cumulative distribution of
$D_{\rm proj}/D$, which characterises the angular distribution (bottom panel) for 
the all three models as labelled. The purple dotted
curve has the form $1 - \sqrt{1 - (D_{\rm proj}/D)^{2}}$, which is what one would 
expected for an isotropic distribution of small blobs.  The no-wind
case is closest to having a central blob topology, while winds (particularly
vzw) moves the gas distribution closer to isotropic.
}\label{fig:proj}
\end{figure}

The top panel of Figure~\ref{fig:proj} shows the projected distances,
$D_{\rm proj}$, of DLAs versus their column densities for all three models.
In general, the median distance is around $\sim 2$~kpc (distance units
are physical unless noted) in all models.  There is a weak trend for
higher-column systems to arise closer to galaxies.  Almost all systems
in all cases arise within 10~kpc of a galaxy.  Hence in our simulations,
DLAs generally arise in the extended neutral gas that are present in
and around galaxies at these epochs.

The middle panel shows the cumulative distribution of projected
distances, $P(>D_{\rm proj})$, for each wind model.  The no-wind case shows
a qualitatively different behavior than the wind runs:  It shows somewhat
larger typical distances at small separations, but beyond $\ga 3$~kpc the
nw case shows very few DLAs, while the wind models possess a significant
tail to high projected separations.  Hence about 95\% of the DLAs in the
no wind case have distances smaller than 5~kpc, but outflows substantially
puff out the neutral gas as shown in Figure~\ref{fig:himaps}, so that
in the vzw case 95\% of DLAs are within 10~kpc.  Since the mass loading
factor is larger for small halos in the vzw model than in the cw model,
the momentum-driven wind model is most efficient at puffing out the
neutral gas from the more numerous and gas-rich small systems.

One can quantify the topology of DLA absorption by examining the
ratio of projection distance $D_{proj}$ to real distance $D$, shown
in the bottom panel.  First, let us consider some simple illustrative
cases: If DLAs come from small blobs that are distributed isotropically
from host galaxies, the cumulative distribution should follow the
purple dotted line, $P(<D_{\rm proj}/D) = 1 - \sqrt{1 - (D_{\rm
proj}/D)^{2}}$.  This is likely similar to a scenario in which DLAs
arise from randomly-distributed small disks within a galactic halo,
as forwarded by \citet{mal01}.  Conversely, if all DLAs are located
in the cross-section plane, then the cumulative distribution should
be a step function at a value set by the angle of the plane relative
to the line of sight.  A special case of this is if the DLA owes
to a single, spherical gas blob centered on a galaxy (``central
blob"), in which case the distribution will be a step function
around unity since $D_{\rm proj}\approx D$.

Now let us examine the actual distributions.  The three models generally
lie between isotropically-distributed blob and central blob geometries,
not surprisingly since these are the extreme cases. In more detail, the
no-wind model lies distinctly more towards central blob, because most of
the DLA cross-section is concentrated in the central galaxy.  Conversely,
the wind models follow more closely to the isotropically-distributed blob
case, with the vzw case more so than the cw.  This means that outflows
change the simple centrally-dominated neutral gas topology of the no-wind
model to a more complex and extended gas topology.  This is consistent
with the general impression one gets from examining the \ion{H}{i}
images in Figure~\ref{fig:himaps}.

These topological trends directly correlate with the DLA kinematics
examined in \S\ref{sec:stats}: The more the winds push out gas, the
larger the velocity widths.  The extra velocity can arise owing to
the kinematics of the outflowing gas itself, or else the fact that
the gas velocity field typically differs more from the galaxy
systemic velocity as one goes farther out into the halo; we dissect
these scenarios further in the next section.  In either case, the
effectiveness by which the momentum-driven winds push out gas into
the halo, owing largely to the high mass loading factors in small
galaxies, is critical for yielding its agreement with DLA kinematics.

\subsection{Relations between DLAs, host galaxies, and host dark matter halos}

\subsubsection{The environment producing high-$\Delta v$ DLAs}

\begin{figure}
\vskip-0.1in
\setlength{\epsfxsize}{0.55\textwidth}
\centerline{\epsfbox{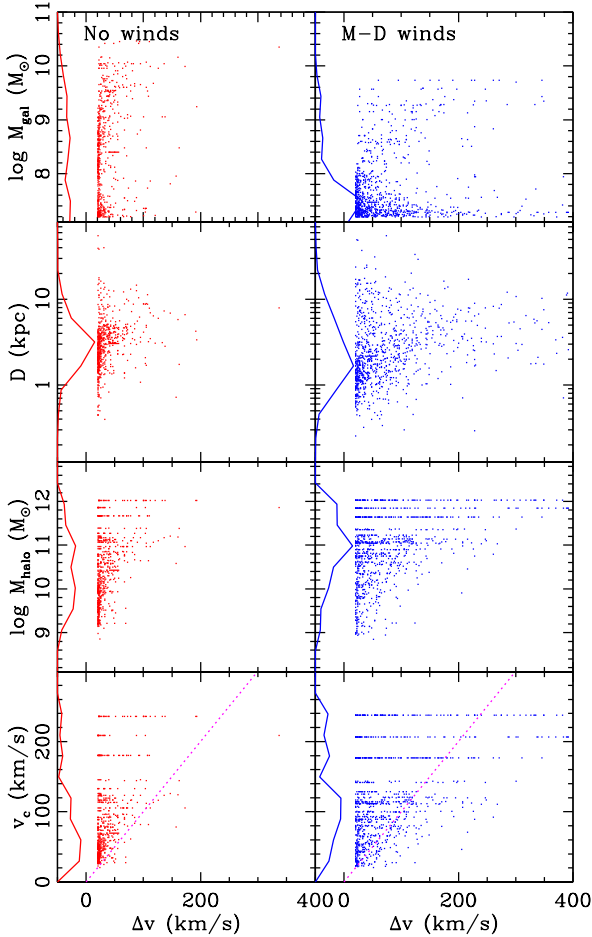}}
\vskip-0.2in
\caption{The velocity width of a DLA, $\Delta v$, versus the masses of the 
nearest galaxy (top panel), 
the distances from the nearest galaxy (second), 
and the mass (third) and circular velocity (bottom) of the host
dark matter halo.  Red points show the nw80 model, blue points
the vzw80 model.
Histograms along the $y$-axis show the distribution of DLAs along
that axis, with a tick mark indicating the median value.
In the bottom panel, the dotted line shows $\Delta v=v_c$; DLAs
to the right of this have ``supergravitational motion", which is
more common in vzw compared to nw.
}\label{fig:20}
\end{figure}

As pushing out the gas distribution into the surrounding halo seems
to be an important aspect of reproducing DLA kinematics, we now
investigate how halo environments impact the kinematics of DLAs.
Figure~\ref{fig:20} shows the relations between the velocity widths
of DLAs, $\Delta v$ and the mass of nearest galaxy $M_{\rm gal}$
(top panel), the distances from the nearest galaxy $D$ (second
panel), the circular velocity of the host dark matter halo $v_c$
(third panel), and the mass of the dark matter halo $M_{\rm
halo}$ (bottom panels) for the no-wind (red points) and the
momentum-driven wind (blue points) models.  The constant wind case
is not shown for clarity, as its trends are intermediate between
the two displayed cases.

The $M_{\rm gal}-\Delta v$ plot shows that $\Delta v$ does not have
a simple relation to the mass of the host galaxy.  The nw model
shows no significant trend, while the vzw model shows an interesting
bimodal distribution for high velocity width DLAs, in which the
nearest galaxies tend to be the most massive or the least massive.
We will see below that these large-$\Delta v$ systems tend to occur
in massive halos, and hence we can deduce that the high-velocity
tail in the vzw model occurs primarily around either massive (likely
central) galaxies or low-mass galaxies within a massive halo.
Overall, more of the absorption seems to come from low-mass galaxies
in the vzw case, showing that the high mass loading factor in small
systems produces significantly more DLA-absorbing cross section
from these galaxies.

The $D-\Delta v$ plot shows that the highest velocity widths occur
relatively far out from the host galaxy, typically $2-10$~kpc away.
The overall median distance is approximately 2~kpc for the vzw case and
3~kpc for nw (shown as the tick marks above the vertical histograms),
but most of the DLAs with $\Delta v>150$~km/s lie above these values.
This means that if one wants to obtain large $\Delta v$ values, one has
to arrange a significant amount of DLA cross-section at relatively large
distances from the host galaxy.  At smaller distances, the internal
kinematics of galaxies is insufficient to produce large velocity
widths.  The vzw model, relative to the nw case, clearly produces more
DLA-absorbing gas distributed farther away from galaxies.

The $M_{\rm halo}-\Delta v$ plot (third panel) shows an envelope
of increasing $\Delta v$ in larger halos.  In the no-wind case, the
envelope occurs at roughly an equality between $v_c$ and $\Delta
v$, showing that with only gravitationally-induced motion, it is
rare to produce kinematics larger than the halo circular velocity.
Conversely, the vzw case shows significant numbers of DLAs with
widths greater than $v_c$.  This is of course key to producing large
velocity widths, and we will explore this further in the next
section.  Note that a massive halo is a necessary but not sufficient
condition for large velocity widths, as massive halos can host DLAs
with small $\Delta v$.  These trends are mirrored in the circular
velocity panel, which encodes information about the halo radius but
otherwise shows very similar trends.

From the above analysis, along with visual impressions from
Figure~\ref{fig:himaps}, we can assemble a scenario for DLA absorption:
Small velocity width DLAs can be produced in all halos, but large velocity
widths require a massive halo environment.  In the momentum-driven
wind scenario, this environment is one where a massive central galaxy
is surrounded by numerous small galaxies in a clustered environment.
The DLA then occurs at intermediate distances away from the nearest
galaxy, which can often be the largest galaxy or one of the smaller
systems.  The wide DLA velocity widths are then produced either by
sightlines intersecting multiple galaxies within a dense region of
protogalactic clumps~\citep[HSR98;][]{mal01}, or by dense gas puffed up
by outflows.  The former can occur without winds, but the latter requires
momentum-driven outflows that produce a puffier neutral gas distribution
and raise the \ion{H}{i} cross-section of small galaxies within larger
halos, thereby yielding the correct fraction of wide systems.

\subsubsection{Supergravitational motion}

In the no-wind model there is no kinetic energy injection into the
neutral gas beyond that provided by gravity.  High velocity width
DLAs with $\Delta v>v_c$ can still arise owing to merging or infalling
motion of neutral gas, as in HSR98, but this is rare.  Conversely,
outflows provide substantial non-gravitational kinetic energy, which
is reflected in the significantly larger fraction of DLAs that have
$\Delta v>v_c$ as seen in Figure~\ref{fig:20} (fourth panel).  We
call the kinematics induced by non-gravitational energy
``supergravitational motion".  The amount of supergravitational
motion ($\Delta v>v_c$) therefore quantifies the contribution of
wind feedback to the kinematics of the neutral gas.

\begin{figure}
\vskip -0.4in
\setlength{\epsfxsize}{0.45\textwidth}
\centerline{\rotatebox{270}{\epsfbox{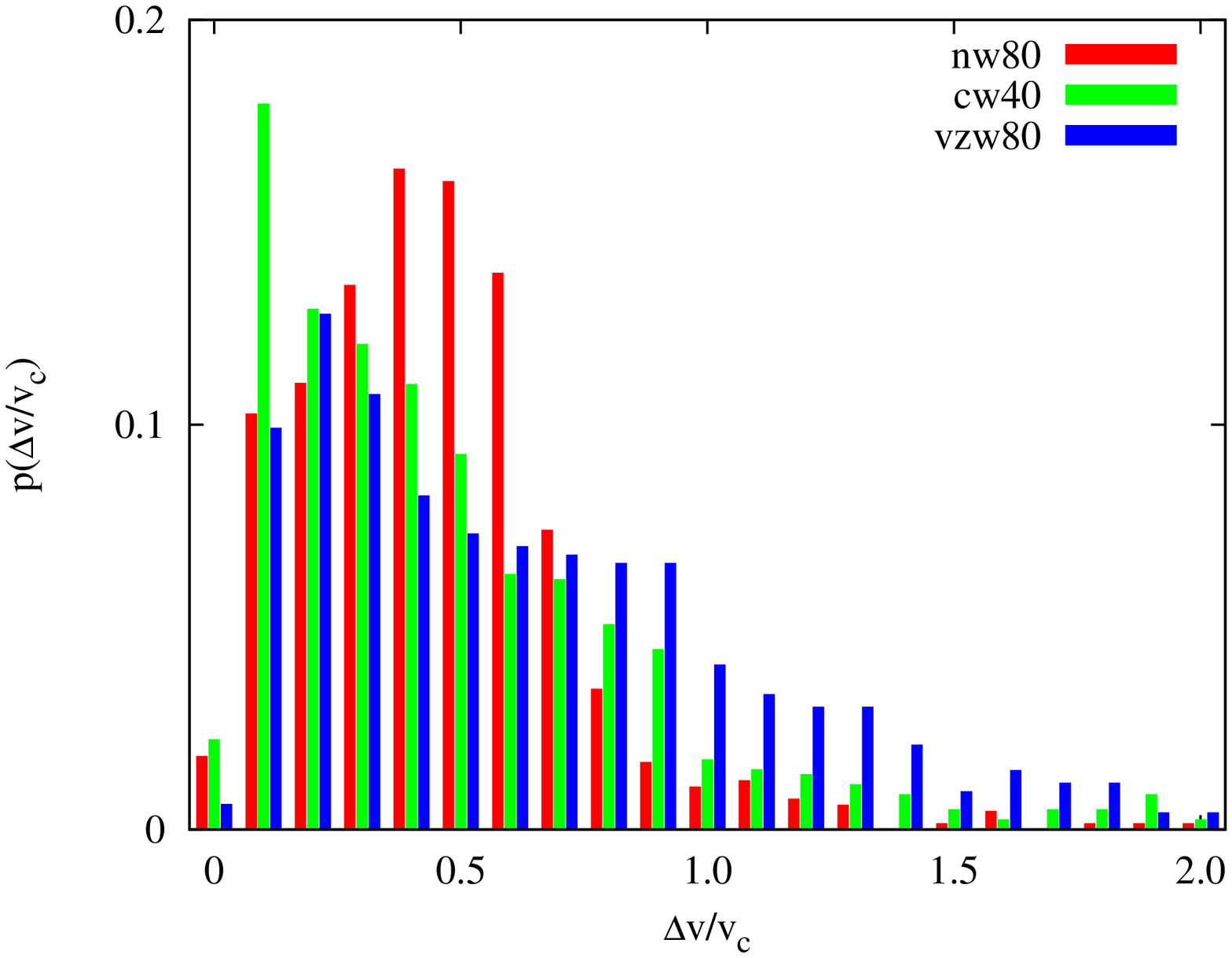}}}
\vskip -0.1in
\caption{The probability distribution $p(\Delta v/v_c)$ for each wind model. 
The fraction showing supergravitational motion,
i.e.$\Delta v/v_c > 1$, are 5\%, 13\%, 
and 22\% for the nw model, the cw model, and the vzw model, respectively.
}\label{fig:dvvc}
\end{figure}

\begin{figure*}
\vskip -0.3in
\setlength{\epsfxsize}{0.9\textwidth}
\centerline{\epsfbox{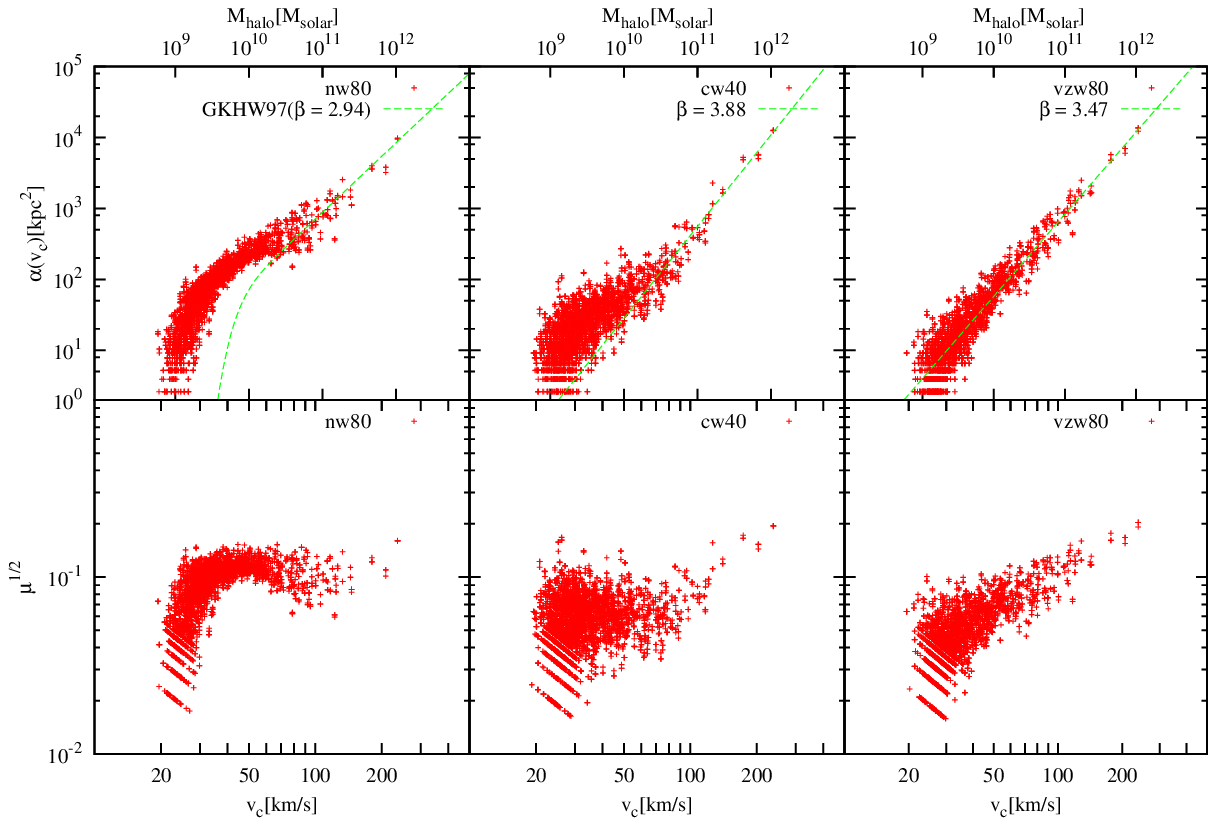}}
\caption{The DLA cross-section $\alpha$ (top panels) and the ratio
$\mu \equiv \alpha/\pi R_{vir}^2$ (bottom) as a function of circular
velocity $v_c$ for the three wind models.  Top axis is labeled by
the approximate halo mass.  Dashed lines in top panels indicate
power-law fits to $v-c>80$~km/s cross sections.  nw80 is consistent
with \citet{gar97a} for large halos, and cw40 is consistent with
Nagamine et al. (2004); in both cases, there is extra absorption
at low $v_c$ relative to an extrapolation from high-$v_c$.  vzw80,
in contrast, shows a pure power law with index $\beta=3.47$ down
to the smallest halos. The other two models have a larger cross-section
for small halos, which is one reason why they produce too many
low-$\Delta v$ DLAs.
}\label{fig:xsection}
\end{figure*}

Figure \ref{fig:dvvc} shows the probability distribution $p(\Delta
v/v_c)$ for each model.  The probability distribution for the no-wind
case is consistent with previous work~\citep{hae00}.  The peak of
$\Delta v/v_c$ is around 0.5 and the fraction of systems with
supergravitational motion is only 5\%.  The no-wind case shows a
strong maximum at $v/v_c\approx 0.5$, and rapidly drops off to
higher $v/v_c$.  The wind models show a qualitatively different
behaviour, with a less obvious maximum and a longer tail to high
$v/v_c$.  The fraction with supergravitational motion for the cw
model and the vzw model are 13\% and 22\%, respectively.  Without
such high fractions of supergravitational motion, the models cannot
reproduce the incidence of wide DLA velocity widths.  Hence the
distribution of gas by outflows, together with small-scale galaxy
clustering, are critical for explaining DLA kinematics.

\subsubsection{DLA cross-section of dark matter halos}\label{sec:crosssection}

To quantify the increase in the puffiness of the gas distribution in
the wind models that is responsible for the wider velocity widths,
we calculate the DLA cross-sections for each dark matter halo.

The upper panels of Figure \ref{fig:xsection} show DLA cross-sections
$\alpha$ in comoving kpc$^2$ for the nw80, cw40, and vzw80 models.
We calculate the cross-section from the projected neutral hydrogen maps,
including only those pixels that produced DLAs with $\Delta v>20$~km/s.
Each DLA produces 3 points on this plot, for projections along each of the
cardinal axes.  The relation between $\alpha$ and $v_c$ has traditionally
been characterized as a power law with a slope we call $\beta$, though
as is evident from the plot a pure power-law is only a good description
for the vzw80 model (upper right).  We note that if the cross-section
was proportional to halo mass, then $\beta=3$, since $M_h\propto
v_c^3$~\citep[e.g.][]{mo98}; this turns out to be very roughly the case,
though the departures from this are key for understanding DLA kinematics.

The no-wind case shows a relatively shallow power law (i.e. low
$\beta$) that results in large contributions to DLA cross-section
from small $v_c$ systems.  This causes a mean velocity width that
is too small when the overall DLA abundance is matched.  For
comparison, we show the fit from the models of \citet{gar97a}, which
also did not include winds but had much lower spatial resolution.
Gardner et al. matched the observed DLA abundance, but in fact the
cross-section in small systems was underpredicted compared to our
current nw model.  Hence, using higher resolution simulations without
winds tend to exacerbate the disagreement with observed DLA kinematics.

For the constant wind model, the power law index is $\beta=3.88$
for $v_c>80$~km/s.  This wind model is identical to that used in
\citet{nag04}, but our slope is slightly larger than theirs because we
fit the slope for $v_c>80$~km/s only.  We choose this cutoff to highlight
where the power law breaks in the cw model, especially in relation to the
vzw case.  For smaller values of $v_c$, the behaviour of the cw model
tends towards that of the no-wind case.  Hence the constant-wind case
is able to puff out gas to some extent, but still produces significant
absorption in low-$v_c$ systems that causes it to match the observed
kinematics less well.

Unlike the other two models, the vzw model shows no break in $v_c$
vs. cross section, with an overall slope of $\beta=3.47$.  This low
cross-section at small $v_c$ is critical for reproducing the observed
kinematics.  As argued in PW97 and \citet{pro01}, most CDM galaxy
formation models fail to reproduce DLA kinematics because they
predict too many small baryonic structures.  The key aspect of the
vzw model is that it has a high mass loading factor $\eta$ in small
galaxies, which lowers the cross-section of the smaller systems by
ejecting more of its material.  Indeed, the reduction in cross
section is roughly equal to $(1+\eta$) in smaller halos as expected
from simple gas-removal arguments: cw is lower than nw by $\sim\times
3$, and vzw is increasingly suppressed to smaller systems since
$\eta\propto v_c^{-1}$.

The lower panels of Figure~\ref{fig:xsection} show $\mu = \alpha/(\pi
R_{\rm vir}^2)$, i.e. the ratio of the DLA absorption cross-section
to the projected area of the dark halo.  For the no wind case $\mu$
is almost constant and cuts off rapidly for the smallest halos, and
the cw case shows qualitatively similar behavior, lowered by
$\sim\times 3$ and with somewhat more scatter.  In contrast, in the
vzw wind model $\mu$ monotonically increases with halo circular
velocity, and hence there is relatively more DLA cross-section at
larger circular velocities that can give rise to high $\Delta v$
values.

\begin{figure}
\vskip -0.3in
\setlength{\epsfxsize}{0.5\textwidth}
\centerline{\rotatebox{270}{\epsfbox{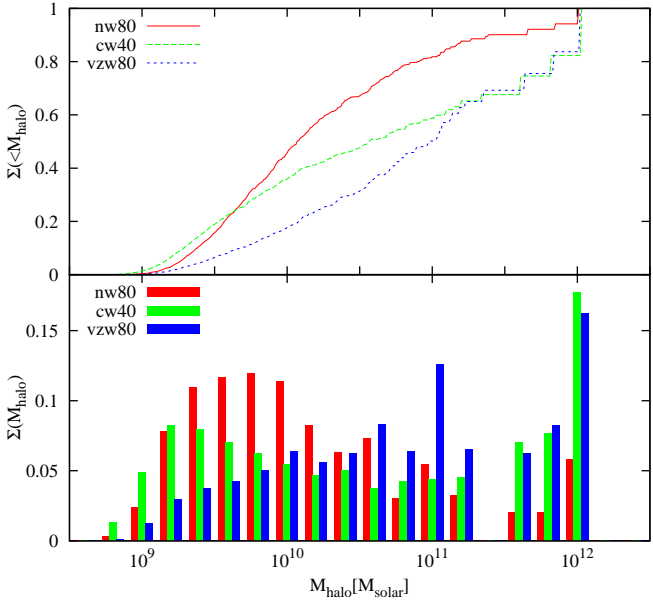}}}
\caption{The cumulative global DLA cross-section (top) and
the differential global DLA cross-section (bottom).
}\label{fig:totxsection}
\end{figure}

In Figure \ref{fig:totxsection} we examine the global cross-sections
to quantify which halo masses are predominantly responsible for DLA
absorption, plotted cumulatively (top) and differentially (bottom) versus
halo mass.  We define the cumulative global cross section to be
\begin{equation}
\Sigma(<M_{halo}) = \int^{M_{halo}}_{0} \alpha(M_h) n(M_h) dM_h
\end{equation}

The cumulative global cross-section reiterates the result that the
vzw model is very efficient at suppressing DLA absorption in small
halos relative to larger ones.  The cw model produces comparably
high cross-sections in large halos to vzw, but it does not suppress
the low-mass halo cross-section, and instead its suppression occurs
more at intermediate masses.

In summary, the combination of both efficiently suppressing absorption
at low masses and having high absorption at large masses relative
to the no-wind case is what makes the momentum-driven wind model
successful at matching the DLA kinematics.  The constant wind case
produces the latter requirement but less so the former, making it
intermediate between the vzw and no-wind cases.

\section{Conclusions}

We examine the kinematic and physical properties of DLAs using
cosmological hydrodynamic simulations including three different
heuristic prescriptions for how galactic outflow properties are
related to their host galaxies: A model without outflows (nw), a
constant wind (cw) model where the wind speed of 484~km/s and mass
loading factor of 2 are independent of galaxy properties, and a
model with momentum-driven wind scalings (vzw) where the wind speed
scales as the circular velocity and the mass loading scales inversely
with it.  Our no-wind model shows results consistent with previous
studies of hierarchical models without strong
outflows~\citep[e.g.][]{pon08}, and dramatically fails to explain
the observed kinematics of DLAs, particularly their systemic velocity
widths as traced by low-ionization metal lines.  This contrasts
with HSR98 who concluded that merging of protogalactic clumps
(without winds) can explain the kinematics of DLAs; in our case,
the high frequency of wide lines cannot be reproduced when such a
model is placed in a cosmological context.  Hence without
non-gravitational motion, it appears that DLA kinematics are not
reproducible within modern hierarchical structure formation models.

We then explored DLA kinematics including outflows.  The central result
of this paper is that our momentum-driven wind model provides a very good
match to all observed DLA kinematic measures, with the possible exception
of the 2-peak test that no model matches well.  Meanwhile, a model with
constant wind speeds and mass loading factors~\citep[as in][]{nag07}
is only marginally consistent with data, and a no-wind case is in strong
disagreement with data, particularly in the distribution of DLA velocity
widths ($\Delta v$) that have historically been difficult to reproduce
in CDM-based models.  This shows that galactic outflows can reconcile
DLA kinematics within current galaxy formation models with observations.
Moreover, DLA kinematics could potentially provide interesting constraints
on the properties of galactic outflows at high redshifts.

The abundances of DLAs may also provide interesting constraints.  Our
no-wind case could not reproduce the observed abundances without a very
high and likely unphysical threshold density for self-shielding.
Meanwhile, both our wind models can match the abundance of DLAs
with reasonable choices for this parameter.  The previous study of
\citet{gar97a} with no winds matched the observed abundance by an incorrect
extrapolation of the DLA cross-section versus circular velocity from
low resolution simulations.  Our higher-resolution simulations show
that such an extrapolation is not valid in the the no-wind or (to a
lesser extent) the constant wind case, since these models produce excess
cross-section at low-$v_c$ over such an extrapolation.  Still, given the
crudeness of our self-shielding criterion, it is difficult to assess the
robustness of this result.  We note that the sophisticated radiative
transfer simulations of \citet{pon08} which include self-consistently
driven outflows are able to match DLA number densities (as well as
metallicities), but they did not match the kinematics.

To understand why outflows help with DLA kinematics, we investigated
the properties of DLA absorption as a function of galaxy and halo
properties.  A consistent story emerges that momentum-driven winds
most efficiently expel gas into a more extended, clumpy distribution
around galaxies, particularly for small galaxies living in dense
environments.  As a result, the DLA cross-section becomes weighted
away from small halos (despite arising preferentially in small
galaxies) towards more massive halos, where forming structures and
outflow-induced motions can yield higher velocities.  This results
in substantial amounts of super-gravitational motion that directly
translates into wide DLA systems.  

HSR98 claimed agreement with observed DLA kinematics by noting that
their individual DLA simulation produced a velocity width of 60\%
of the virial velocity $v_{\rm vir}$, and then integrating over the
halo mass function assuming that the DLA cross-section scales as
$v_{\rm vir}^3$.  In our comparable no-wind simulation, this latter
scaling is appropriate for larger halos, but it becomes much shallower
for smaller ones, yielding much more DLA absorption at small cross
sections.  This results in poor agreement with the observed kinematics
in the no-wind case.  In another study, \citet{mal01} found that
gaseous disks of standard sizes were unable to reproduce DLA
kinematics, and that more extended gaseous distributions were
required.  While they mostly focused on testing extended disks
models, they noted that outflows could potentially have a similar
effect.  Our momentum-driven wind case appears to produce just the
required gas distribution.  We note that disk sizes are generally
too small without strong feedback and much higher numerical resolution
than we have here, but the results of \citet{mal01} suggest that
even optimally conserving angular momentum is not enough, and some
supergravitational motion is required.  Finally, while outflows
produce greater velocity widths, they also produce less cross-section
overall.  We hypothesize that such models will therefore produce
more Lyman Limit Systems (LLS; $10^{17}<\nhi< 10^{20.3}\cdunits$);
\citet{dav10} showed that winds do produce more strong \lya\ forest
systems ($\nhi>10^{14}\cdunits$) at low-$z$, but did not have
sufficient statistics to explore LLS.  CDM models have traditionally
also had difficulty producing enough LLS \citep[e.g.][]{her96},
though the strength of the discrepancy has been difficult to quantify
owing to uncertainties in measuring column densities in this regime.

More broadly, we note that momentum-driven winds have enjoyed
substantial success in reproducing a wide range of observations of
early galaxies and intergalactic gas, particularly related to cosmic
chemical evolution.  On the other hand, this investigation of DLAs
focuses solely on the neutral gas; metals are merely used as a
tracer of dense \ion{H}{i}.  Combining these various results leads
us to conclude that outflows must move not only metals but also
substantial amounts of {\it mass} around the cosmos.  It may be
possible to construct a model that enriches the IGM by preferentially
expelling highly enriched gas from galaxies which could still
reproduce the metal-line kinematics, but the strong low-ionization
metal absorption generally indicates a substantial neutral hydrogen
component, if for nothing else than to shield it from being sent
into higher ionization states.  Furthermore, the high mass outflow
rates~\citep[typically comparable to or greater than the mass forming into
stars;][]{opp08} suggest that a large amount of ISM gas must be
entrained in these outflows, and hence the outflow metallicity
cannot be substantially greater than the ambient ISM.  Hence it is
difficult to envision a scenario where the metals are being ejected
highly preferentially compared to the mass.  The fact that a single
outflow model where vast amounts of gas are expelled at the typical
ISM metallicity reproduces all these observations within a cosmological
context is quite remarkable.  Nevertheless, the lack of possibly
important physics in our current simulations precludes any definitive
constraints on outflows from DLA kinematics.

DLA kinematics have long been an oddity that did not seem to fit neatly
into our current hierarchical view of galaxy formation.  Our study
suggests that the answer is not in modifications to the hierarchical
view, but rather to modifications in the detailed processes of how
galaxies form.  In a sense, this is a more exciting possibility, as it
allows us to move towards constraining such processes from a completely
new and different perspective.  The work presented here paints a hopeful
picture that DLA kinematics may fit into a broader understanding of
galactic outflows.  However, our current form of modeling outflows is
relatively crude, and the present analysis did not address the wider
range of DLA observations such as metallicities and redshift evolution
that could provide interesting constraints.  Furthermore, we did not
include detailed radiative transfer to more accurately predict the
distribution of neutral gas.  Hence this work must be considered as a
preliminary step that merely highlights the importance of a new physical
process in yielding DLA properties as observed; we plan to examine a
more comprehensive suite of DLA properties in the future.  This will
help us obtain a more complete picture for how DLAs relate to the
evolution of galaxies in a hierarchical universe.

\end{document}